\documentclass[11pt]{article}

\usepackage{nicefrac}
\usepackage{fancyhdr}

\begin{document}

\textheight23cm
\topmargin-0.5cm
\textwidth16cm

\newcommand{\bin}[2]{\left(\begin{array}{c} \!\!#1\!\! \\  \!\!#2\!\!
\end{array}\right)}

\huge
\begin{center}
Random-matrix theory for complex atomic spectra
\end{center}

\normalsize

\vspace{0.5cm}

\Large

\begin{center}
Jean-Christophe Pain
\end{center}

\normalsize

\begin{center}
CEA, DAM, DIF, F-91297 Arpajon, France

jean-christophe.pain@cea.fr
\end{center}

%===============
%	ABSTRACT
%===============

\begin{abstract}
Around 1950, Wigner introduced the idea of modelling physical reality with an ensemble of random matrices while studying the energy levels of heavy atomic nuclei. Since then, the field of random-matrix theory has grown tremendously, with applications ranging from fluctuations on the economic markets to complex atomic spectra. The purpose of this short article is to review several attempts to apply the basic concepts of random-matrix theory to the structure and radiative transitions of atoms and ions, using the random matrices originally introduced by Wigner in the framework of the gaussian orthogonal ensemble. Some intrinsic properties of complex-atom physics, which could be enlightened by random-matrix theory, are presented.
\end{abstract}

\normalsize

\vspace{0.5cm}

{\bf Keywords}: Atomic spectra, random-matrix models, spectral lines. 

{\bf PACS}: 31.15.bt, 32.70.Cs, 02.50.Cw, 02.10.Yn

\vspace{1cm}

%===================
%	INTRODUCTION
%===================

\section{Introduction}

The structural and radiative properties of multi-charged ions are crucial for the studies of laboratory and astrophysical hot plasmas. Detailed numerical calculations, however, are complicated and even sometimes impossible. Indeed, when the thermodynamical conditions are such that several shells are open, the total number of electronic configurations and electric-dipole lines can become huge. In that case the computation becomes numerically untractable. Complex atomic spectra \cite{JUDD85} may be characterized by statistical (global) methods, particularly if one is interested only in general properties and regularities. The atomic spectrum as a distribution of energy levels or spectral lines can for instance be described by its statistical moments, which define its center of gravity, width, asymmetry (skewness), sharpness (kurtosis) and other mean characteristics \cite{KENDALL69}. A limited number of moments enables one to approximately describe the complex spectra with unresolved transition arrays \cite{HARRISON31}, to reveal the statistical properties of spectra, to analyze the relative influence of various interactions, to study changes in atomic or isoelectronic sequences, to investigate configuration mixing, \emph{etc}. The first problem is to find explicit expressions of such average characteristics \cite{BAUCHE79,KARAZIJA91}, without any detailed calculation of the spectrum. The second problem is the choice of the modelling function \cite{GILLERON08,PAIN09}. Having explicit formulas for a number of first moments enables one to approximately restore the envelope of the radiation spectrum. Studies of variation of these statistical characteristics along isoelectronic sequences give a wealth of information on intra-atomic interactions \cite{RUDZIKAS97}. 

However, such a statistical approach does not provide any information about the internal correlation laws of a transition array. It gives no indication about the possible occurence of chaotic phenomena.

The random-matrix theory (RMT) \cite{MEHTA67,BRODY81,CAMARDA83,KUNZ98} is a stochastic approach which relies on the assumption that the Hamiltonian matrix represents an ensemble of real finite matrices characterized by the values of their elements, which are defined by their differential probability to occur. The matrix probability distribution is assumed to consist of a product of distributions for the individual matrix elements (\emph{i.e.} the matrix elements are statistically independent). Wigner's gaussian orthogonal ensemble (GOE) consists of real symmetric matrices where the distribution of the diagonal elements is given by a Gaussian with a zero mean and a variance twice that of the off-diagonal elements. Obviously, the matrix elements of the real system are not distributed independently. However, the validity of the approximation can be tested on the predictions made for the eigenvalue spectrum. RMT predicts the way adjacent levels are correlated (statistical ``repulsion'' of levels).

The successful application of RMT to the study of the distribution of energy levels of atoms and ions is recalled in section \ref{sec2}, and its extension to the calculation of electric-dipole (E1) lines in section \ref{sec3}.  An interesting alternative, the Two-Body Random Matrix Model is also explained in section \ref{sec3}. In section \ref{sec4}, several unexplained properties of complex atomic spectra are presented, within an attempt to relate them to RMT. Section \ref{sec5} is the conclusion.

%====================
%	ENERGY LEVELS
%====================

\section{\label{sec2} Energy levels}

In the atomic central-field model, the Hamiltonian is diagonalized in electronic configurations. The distribution of the matrix elements is so complex that it can be considered as random. In the RMT, all the elements of a real symmetric matrix in Wigner's GOE are independent random variables obeying a Gaussian distribution. The eigenvalues of a random matrix are highly correlated; for instance, two adjacent eigenvalues ``repel'' each other. The distribution of the eigenvalues themselves tends to become Gaussian when the number of interacting particles increases. In contrast, numbers chosen at random are not correlated. More precisely, if the level spacings are considered as a set of numbers $x_{i+1}=x_i+s$ with $x_1=0$ such that $s$ is a random variable \cite{BAUCHE90}, their distribution obeys Poisson's law:

\begin{equation}
p\,(s)\; ds=e^{-s}\; ds.
\end{equation}

With random correlations of the matrix elements $s$ of $H$, the level spacings follow the Wigner surmise

\begin{equation}\label{wig}
p\,(s)\;ds=\frac{\pi}{2}\;s\;e^{-\pi s^2/4}\;ds,
\end{equation}

which holds in the GOE. Actually, the GOE represents a special case of a large family of ensembles which all have the same spectral behaviour. 

Rosenzweig and Porter \cite{ROSENZWEIG60} plotted the empirical distribution of nearest-neighbor spacings for the odd-parity levels of neutral Hf ($Z$=72). The Wigner distribution (see equation (\ref{wig})) was found to give an excellent qualitative agreement with the empirical distribution. Later, Flambaum \emph{et al.} \cite{FLAMBAUM94} studied the spacing distributions of the $J^{\pi}=4^+$ levels of Ce ($Z$=58). They found that the experimental level spacings were in good agreement with the Wigner distribution. O'Sullivan \emph{et al.} \cite{OSULLIVAN99} pointed out that taking configuration interaction into account is crucial in order to obtain a Wigner distribution for the level spacings. The single-electron states mix together and produce a spectrum of levels analogous to the one in compound nuclei. In fact, if the ratio of the mean off-diagonal Coulomb interaction parameter by the mean level spacing is large, the single-particle model no longer holds and the states highly mix so that only total angular momentum $J$ and parity $\pi$ are good quantum numbers.

Studying the distribution of the adjacent-level spacings enables one to check how far the level density lies from the Poisson and Wigner (derivative of a Gaussian) limits \cite{CASATI90,FEINGOLD91}. In the GOE, a $\chi^2$ distribution can be derived from the Porter-Thomas law \cite{PORTER56,PORTER65} for the strength of the lines (see section \ref{subsec43}).

Bohigas, Giannoni and Schmidt \cite{BOHIGAS84} conjectured that the spectrum of a chaotic system is strongly related to the spectrum of a random matrix. More precisely, they found that the level fluctuations of the quantum Sinai's billiard are consistent with the predictions of the GOE, which was a confirmation that RMT is strongly linked to quantum chaos and that the fluctuation laws are universal. In fact, for classical systems, non-integrable processes follow Wigner distribution, while the properties of integrable systems obey Poissonian statistics. The extrapolation from classical chaotic systems to quantum ones was discussed for instance by Connerade \emph{et al.} \cite{CONNERADE97,CONNERADE98}. Yukawa recovered the GOE predictions without introducing \emph{a priori} ensembles of random matrices. By means of the usual statistical laws of classical many-body systems, he proposed a level distribution function which makes a transition from the Poisson type to the Gaussian type depending on the value of a single parameter characteristic of the Hamiltonian. 

%=============================
%	ELECTRIC-DIPOLE LINES
%=============================

\section{\label{sec3} Electric-dipole lines}

While the position of the absorption lines depends on the eigenvalues, their strength is mostly governed by the eigenvectors. In an atomic-structure code, most of the time is consumed in diagonalizing the Hamiltonian matrices and constructing the dipole transition matrix. Most of the calculational effort resides in constructing the off-diagonal matrix elements of the blocked Hamiltonian matrices. This is easily understood as these matrix elements are formed not only from one-body operators (as the transition matrix elements) but also from two-body Coulomb interaction operators.

\subsection{\label{subsec31} The RMT as a tool for simplifying the exact calculation}

In their work \cite{WILSON88}, Wilson \emph{et al.} calculated the diagonal terms of the Hamiltonian matrix in a pure coupling using Cowan's atomic-structure code \cite{COWAN81} and populated off-diagonal elements statistically beyond the GOE according to a bi-Gaussian distribution function. They observed a disproportionately large number of off-diagonal elements of small amplitude and noticed that the larger matrix  elements connect basis states where the parent shell is also of common spectroscopic term. A detailed analysis showed that the larger elements are distributed according to the GOE predictions, while the smaller are distributed like a Gaussian with a much smaller width (bi-Gaussian surmise).

\subsection{\label{subsec32} The two-body random interaction model}

The RMT does not describe many important properties of realistic many-body systems, its predictions being limited to level statistics, localization properties of the eigenstates, \emph{etc.} The Two-Body Random Interaction Model (TBRIM) \cite{FRENCH70,BOHIGAS71a,BOHIGAS71b,BOHIGAS75} was introduced as an intermediate approach based on a simple mathematical model with random interactions which, however, takes into account the most important features of many-body systems: orbitals, two-body interactions and the Pauli exclusion principle. It consists of a system of fermions ($n$ fermions distributed in $m$ orbitals with energies $\epsilon_i$ distributed between 0 and 1, see Refs. \cite{FLAMBAUM96,FLAMBAUM97,JACQUOD97,GEORGOT97}). Then, all possible states with $n$ fermions are built, the state $k$ being characterized by $n_i^{(k)}=0$ or $1$, with $i=1,\cdots, m$. The Hamiltonian matrix is such that

\begin{equation}
H_{kl}=\sum_{i=1}^mn_i^{(k)}\epsilon_i~\delta_{kl}+V_{kl},
\end{equation}

and $V_{kl}$ is distributed randomly in the interval $\left[-V_0,V_0\right]$ and is set equal to 0 if $k$ and $l$ states differ by more than two orbitals. $\delta_{kl}$ is Kronecker's symbol. The density of states is found to be a Gaussian:

\begin{equation}
\rho(E)=\frac{1}{\sqrt{2\pi}\sigma}e^{-(E-E_c)^2/2\sigma^2},
\end{equation}

centered in

\begin{equation}
E_c=\frac{1}{N}\sum_{k=1}^NH_{kk}
\end{equation}

and having a variance

\begin{equation}
\sigma^2=\frac{1}{N}\sum_{k=1}^NH_{kk}^2-E_c^2+\frac{1}{N}\sum_{k=1}^N\langle\Delta E^2_k\rangle,
\end{equation}

where

\begin{equation}
\langle\Delta E^2_k\rangle=\sum_{p=1, p\ne k}^NH_{kp}^2.
\end{equation}

The number $N$ is the size of the two-body random interaction matrix:

\begin{equation}
N=\bin{m}{n}=\frac{m!}{n!(m-n)!}.
\end{equation}

and the sparsity $r$ of the matrix is given by \cite{FLAMBAUM98}:

\begin{equation}
r=\frac{1}{N}[1+n(m-n)+n(n-1)(m-n)(m-n-1)/4].
\end{equation}

The diagonalization of the Hamiltonian matrix leads to the following expression of the wavefunctions:

\begin{equation}
|\psi^{(i)}\rangle=\sum_{j=1}^NC_j^{(i)}|\phi_j\rangle,
\end{equation}

where the coefficients $C_j^{(i)}$, defined by

\begin{equation}
C_j^{(i)}=\frac{1}{\rho(E_j)}\frac{\gamma_j}{\pi}\frac{1}{(E-E_j)^2+\gamma_j^2},
\end{equation}

are the components of the eigenstates in terms of the $|\phi_j\rangle$ basis. The $E_j=H_{jj}$ expectation value represents the energy of the many-electron atom state $|\phi_j\rangle$ and

\begin{equation}
\gamma_j=\langle\Delta E^2_j\rangle=\sum_{i=1}^N\left|C_j^{(i)}\right|^2\left(E_j-E^{(i)}\right)^2=\sum_{p=1, p\ne j}^NH_{jp}^2,
\end{equation}

where $E^{(i)}$ is the energy of state $|\psi^{(i)}\rangle$. The configuration interaction leads to the ``straightening'' of the spectrum. This effect is a direct result of the level repulsion. It strongly manifests in the parts of the spectra where the average level spacing is much smaller than the typical matrix element $H_{ij}$ ($i\ne j$).
In complex atomic spectra, if a given interaction mixes efficiently the basis ``natural'' states, the states of the system are composite states with the fractional occupation numbers for the orbitals

\begin{equation}
n_s=\langle\psi^{(i)}|a_s^+a_s|\psi^{(i)}\rangle=\sum_{k=1}^N\left|C_k^{(i)}\right|^2\langle\phi_k|a_s^+a_s|\phi_k\rangle,
\end{equation}

where $a^+$ and $a$ represent the second-quantization creation and annihilation operators respectively. Angom \emph{et al.} \cite{ANGOM03} showed, through a statistical analysis of Sm I eigenvalues and eigenfunctions for $J=0^+$ and $J=4^+$, that atoms with a sufficiently large number of active electrons exhibit TBRIM characteristics. The density of states predicted by the TBRIM is Gaussian and a Gaussian-like density of states was also observed in the Sm IX spectra by O'Sullivan \emph{et al.} \cite{OSULLIVAN99}. The fact that some quantities can be obtained without diagonalization of the Hamiltonian matrix can be useful for the calculation of transition rates \cite{FLAMBAUM98,FYODOROV98}.

%======================================
%	INTRINSIC PROPERTIES OF SPECTRA
%======================================

\section{\label{sec4} Intrinsic properties of spectra}

In this section we mention a few interesting properties of the transition arrays of E1 lines which are not fully understood yet and for which the RMT could bring some explanations.

\subsection{\label{subsec41} Scars of symmetries}

In transition arrays of complex atomic spectra, some well-defined subsets of lines bear the signatures, called scars, of their vanishing in one or several pure couplings. Symmetries are essential in the approach to $n-$particle systems. In atomic and molecular physics, some current studies deal with complex low-symmetry situations. For example, the high symmetry of the hydrogen atom, when broken by an external field, can lead to quantum chaos. However, it was shown by Delande and Gay \cite{DELANDE87} that, for that system,  symmetry effects can also be detected, as scars in the joint distribution of line wave numbers and oscillator strengths. In their paper, Bauche and Bauche-Arnoult \cite{BAUCHE92} presented some scars of symmetries in the distribution of line intensities in complex atomic spectra. They found that in a particular type of transition array, one can define subsets of lines whose lines are, on the average, definitely weaker or stronger than the others. The fact that they are weaker is trivial when the physical coupling is close to an extreme coupling where they vanish. This was shown numerically, by means of line-by-line calculations, in the cases of two particular transition arrays $d^4\rightarrow d^3p$ and $d^8\rightarrow d^7p$. The authors pointed out that such a phenomenon can persist very far from that coupling, which was rather unexpected. Such scars are the traces of the mathematical structure of the intensity distribution, which remains observable in extremely complex situations. The corresponding scars are different from the ones exhibited by Delande and Gay \cite{DELANDE87}, since they are not produced by an external field; they are observed in the intensity distribution without reference to the line wave numbers and they occur in many-electron atomic systems.

\subsection{\label{subsec42} Fractal nature of atomic spectra}

Learner \cite{LEARNER82} found what he called an ``unexpected law'' related to the number of weak lines in a spectrum. The logarithm $\log(N_k)$ of the number of lines whose intensities lie between $2^kI_0$ and $2^{k+1}I_0$ is a decreasing linear function of $k$. In the example of Learner (spectrum of Fe I), the value of $I_0$ is chosen in such a way that this law holds for $1\leq k\leq 9$ (9 octaves) when about 1500 lines within the wavelength range 290 nm $\leq \lambda\leq$ 550 nm are considered. If $p$ is the slope, one has  $N_k=N_0 \times 10^{-kp}$. The number of lines is divided by $10^p$ when the size of the interval is multiplied by two. A characteristic dimension $d$ is therefore given by

\begin{equation}
\frac{1}{10^p}=2^d,
\end{equation}

\emph{i.e.}

\begin{equation}
d=-\frac{p}{\log_{10}(2)}.
\end{equation}

Learner noticed that $p\approx\log_{10}(2)/2$. In that case, $d\approx -1/2$ and it will be shown in a future work that the Hausdorff fractal dimension associated to Learners's rule is close to $1/3$. The link between fractal behavior and the RMT is not obvious, but definitely exists \cite{HUSSEIN00}. An excess of weak lines is also observed in that case, as in the approach of Ref. \cite{WILSON88}. Bauche \emph{et al.} related this excess to the coupling effects \cite{BAUCHE94} when the physical coupling is close to a pure coupling, and to scars of symmetries in the opposite case. 

\subsection{\label{subsec43} The law of anomalous numbers}

The observation by Newcomb \cite{NEWCOMB81} that the first pages of logarithm books were more used than the last ones, led to the conjecture that the significant digits of many sets of naturally occuring data are not equi-probably distributed, but in a way that favors smaller significant digits. For instance, the first significant digit (\emph{i.e.} the first digit which is non zero) will be 6 more frequently than 7 and the first three significant digits will be 439 more often than 462. Benford provided a probability distribution function for significant digits, which states that the probability that the first significant digit $d_1$ is equal to $k$ is given by \cite{BENFORD38}:

\begin{equation}\label{bl2}
P(d_1=k)=\log_{10}\left(1+\frac{1}{k}\right). 
\end{equation}

It was found recently (see Ref. \cite{PAIN08}) that the distribution of lines in a given transition array follows very well Benford's logarithmic law of significant digits. This indicates that the distribution of digits reflects the symmetry due to the selection rules. If transitions were governed by uncorrelated random processes, each digit would be equi-probable. 

Benford's law is still not fully understood mathematically. However, it applies if the system is governed by multiplicative processes \cite{PIETRONERO01}. Indeed, a random multiplicative process corresponds to an additive process in a logarithmic space. As we have seen before, in Wigner's RMT, the Hamiltonian is defined in the GOE by an ensemble of real symmetric matrices for which the probability distribution is a product of the distributions for the individual matrix elements  $H_{kl}$, considered as stochastic variables, and the variance of the distribution for the diagonal elements is twice the one for the off-diagonal elements. The line strength $S$ is proportional to $|\langle i|\vec{D}|j\rangle|^2$, where $\vec{D}$ is the dipole operator and $|i\rangle$ and $|j\rangle$ are eigenvectors of the Hamiltonian. Therefore, the line strengths involve quantities such as products of $H_{kl}$ and one has

\begin{equation}
\log S=\log\zeta+\log S',
\end{equation}

where $\zeta$ is a stochastic variable and $S$ and $S'$ two different strengths. The central-limit theorem states that the probability distribution that the value of the $n^{th}$ strength is $S$ will be Gaussian with a variance
$\propto n^{1/2}$. In the infinite limit, the distribution will approach a constant $K$, so that one has

\begin{equation}
\int P(\log S)d(\log S)=K\int\frac{dS}{S}.
\end{equation}

The probability $P$ that the first significant digit $d_1$ of $S$ is $k$ in base 10 is given by

\begin{equation}
P(d_1=k)=\int_{k}^{k+1}\frac{dS}{S}\Big/\int_{1}^{10}
\frac{dS}{S}=\log_{10}\Big(1+\frac{1}{k}\Big),
\end{equation}

which is exactly equation (\ref{bl2}). This argument relies on the idea that fluctuations are governed by multiplicative processes involving a
stochastic variable. The matrix elements of the Hamiltonian are correlated stochastic variables and the product of such variables leads to Benford's logarithmic distribution of digits.

Another argument can be invoked to understand the validity of Benford's law for the analysis of complex atomic spectra. As we have seen in section \ref{sec2}, Porter and Thomas have shown that the amplitudes of the lines between all the levels of two random matrices obey a Gaussian distribution \cite{PORTER56,PORTER65}, which is 

\begin{equation}\label{port}
D(S)=\frac{L}{\sqrt{2\pi \langle S\rangle S}}\exp\left[-\frac{S}{2\langle S\rangle}\right],
\end{equation}

where $L$ and $\langle S\rangle$ are respectively the number of lines and the average value of the line strength $S$ (the strength is the square of the amplitude) in a ($J,J'$) set. When numbers are taken from an exponential distribution, they automatically obey Benford's law. Therefore, when the exponential term dominates in $D(S)$, which is often the case except close to the origin (\emph{i.e.} for very weak lines), Benford's law applies.

The RMT contains approximate symmetries, which are not sufficient to describe the vicinity of Russell-Saunders (LS) and jj couplings \cite{BAUCHE92,BAUCHE91}. However, we found that even when the RMT is expected to be inappropriate, \emph{i.e.} close to a pure coupling, the line strengths still fit Benford's law. This is due to the fact that even if the number of weak emerging lines is important (when the term $\frac{1}{\sqrt{S}}$ dominates in the Porter-Thomas law (\ref{port})), the number of decades is sufficiently large so that the weight of those weak lines is negligible in the statistical occurence of digits.

Since Benford's law can be explained in terms of a dynamics governed by multiplicative stochastic processes, the RMT is probably an interesting pathway for the calculation of large atomic-dipole transition arrays and Benford's law can help clarifying the existence of different classes of stochastic Gaussian variables.

%=================
%	CONCLUSION
%=================

\section{\label{sec5} Conclusion}

Random-matrix theory offers a new kind of statistical description of disordered systems. It is concerned with the statistical properties of eigenvalue sequences. Since its inception in the field of nuclear physics, this description was found to apply to the statistical description of the eigenvalues of a quantum chaotic system. Although the level sequences that arise from these different physical systems follow different overall distributions, the fluctuation properties that they exhibit seem to be universal. This universality is an important property of RMT but as with any statistical description, detailed information of a system is lost. The strength and relevance of RMT lie in the fact that the fluctuation properties of a system's  eigenvalue sequence are independent of its detailed structure. Rather, the fluctuation properties of such a sequence depend solely on the system's overall symmetries (which are called ``couplings'' in atomic spectroscopy). RMT can therefore be used to get insight into the symmetry properties of a system by studying the statistical properties of its corresponding eigenvalue sequence. Since the fifties, RMT has grown tremendously in scope, and has even become a field of study in its own right. With this growth, the mathematical tools developed allow ensembles of random matrices to model the structural and radiative properties of complex atoms encountered in multicharged-ion plasmas.

\vspace{0.5cm}

{\bf Acknowledgements}

\vspace{0.5cm}

The author would like to thank O. Peyrusse for stimulating discussions and for suggesting interesting references.

\vspace{0.5cm}

%===================
%	BIBLIOGRAPHY
%===================

\end{document}